\documentclass{PoS}
\usepackage{wrapfig}
\usepackage{amsmath}
\usepackage{cite}

\title{{\footnotesize DESY 18-150,~~DO-TH 18/19} \newline 
NNLO Parton Distributions for the LHC}

\ShortTitle{NNLO PDFs for the LHC}

\author{\speaker{Sergey Alekhin}%
         \thanks{This work was supported in part by Bundesministerium 
f\"ur Bildung und 
                 Forschung (contract 05H15GUCC1). 
}\\ 
II. Institut f\"ur Theoretische Physik, Universit\"at Hamburg,
    Luruper Chaussee 149, D-22761 Hamburg, Germany;\\
        Institute for High Energy Physics,142281 Protvino, Russia\\
        E-mail: \email{sergey.alekhin@desy.de}}

\author{Johannes Bl\"umlein\\ 
        Deutsches Elektronensynchrotron DESY, Platanenallee 6, D--15738 Zeuthen, Germany\\
        E-mail: \email{Johannes.Bluemlein@desy.de}}

\author{Sven-Olaf Moch\\
II. Institut f\"ur Theoretische Physik, Universit\"at Hamburg,
    Luruper Chaussee 149, D-22761 Hamburg, Germany \\      
 E-mail: \email{sven-olaf.moch@desy.de}}

\abstract{
We consider some trends, achievements and a series of remaining problems in the 
precision determination of parton distribution functions. For the description 
of the scaling violations of the deep-inelastic scattering data, forming the key 
ingredients to all PDF fits, a solid theoretical framework is of importance. 
It is provided by the fixed flavor number scheme in describing 
the heavy-quark contributions which is found in good agreement with 
the present experimental data in a very wide range of momentum transfers. 
In this framework also a consistent determination of 
the heavy-quark masses is possible at high precision. The emerging Drell-Yan 
data measured at hadron colliders start to play a crucial role in 
disentangling the quark species, particularly at small and large values of $x$. 
These new inputs demonstrate a good overall consistency with the earlier constraints 
on the PDFs coming from fixed-target experiments. No dramatic change is observed in 
the PDFs in case of a consistent account of the higher-order QCD corrections and 
when leaving enough flexibility in the PDF shape parameterization.}

\FullConference{Loops and Legs in Quantum Field Theory (LL2018)\\
		29 April 2018 - 04 May 2018\\
		St. Goar, Germany}

\begin{document}

After a long period of phenomenological studies, contemporary particle physics
has reached the level of percent accuracy for the parton distribution functions (PDFs). 
However, some important features still need further clarification~\cite{Accardi:2016ndt}. This concerns in particular
the asymptotic behavior for small and large values of Bjorken $x$. The first issue is in 
turn related to the theoretical foundations for the description of small-$x$ deep-inelastic 
scattering (DIS) processes, including the heavy-quark contributions to the structure 
functions due to charm and bottom.
The latter provide very essential constraints on the PDFs in the small-$x$ region. 
The heavy quark contribution to deep-inelastic scattering (DIS) is commonly considered 
within two competitive factorization schemes, one with a fixed number of flavors
(FFN) and another variable number of flavors (VFN). A detailed comparison of these 
two approaches was performed in Ref.~\cite{Accardi:2016ndt} and the FFN scheme 
was found to provide a better description of the existing HERA data on
DIS charm production. Indeed, the superiority of the FFN scheme versus a VFN scheme within the kinematic
region covered by HERA had been observed already very early on, cf.~Ref.~\cite{Gluck:1993dpa}.

\begin{wrapfigure}{l}{0.5\textwidth}
\begin{center}
\includegraphics[width=0.49\textwidth, angle=0]{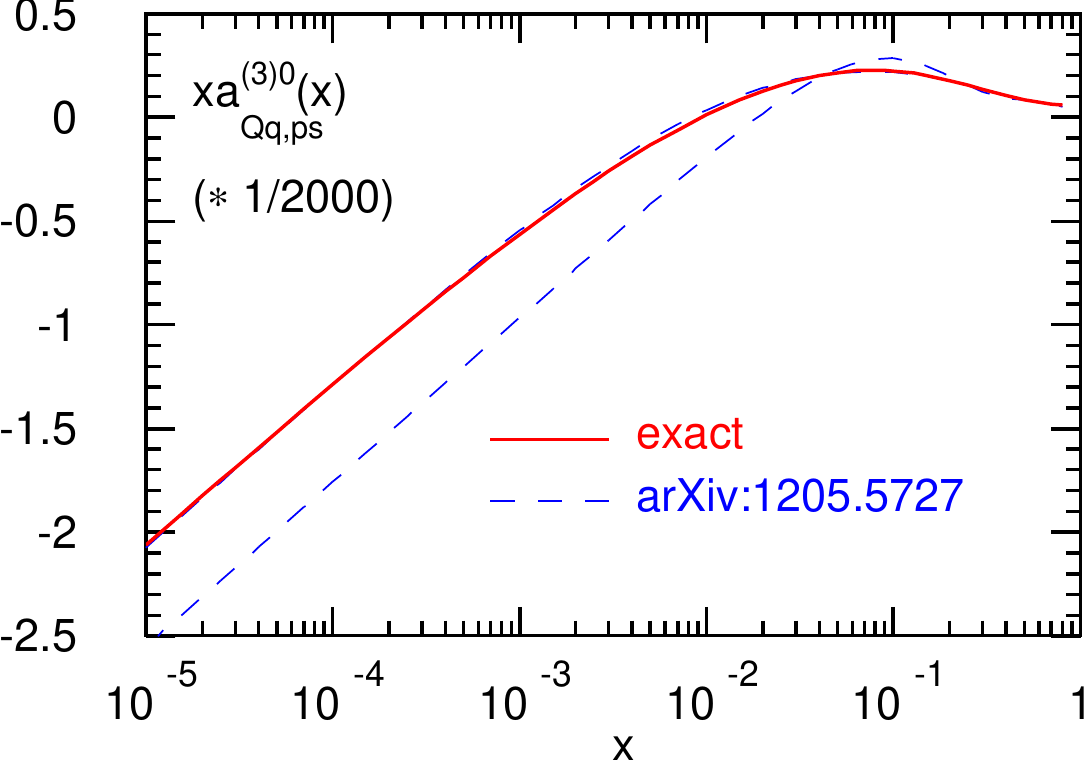}
\caption{\small 
\label{fig:aQj30} 
The exact result for the 3-loop pure single OME $a^{(3)\,0}_{Qq,\,\rm ps}$ \cite{Ablinger:2014nga} 
and the comparison to previous approximations of Ref.~\cite{Kawamura:2012cr}
based on the limited set of Mellin moments from Ref.~\cite{Bierenbaum:2009mv}.
}
\end{center}
\end{wrapfigure}
\noindent
The FFN scheme turns out to provide a more consistent setting for 
the heavy quark masses than in the case of the VFN scheme~\footnote{See also Ref.~\cite{H1:2018flt} 
for an updated comparison with the use of recent HERA data on 
the heavy-quark production}. Present theoretical calculations in the 
FFN scheme include the next-to-next-to-leading order (NNLO) Wilson coefficients~\cite{Kawamura:2012cr}, which are modeled using 
the available asymptotics in different kinematic regimes between heavy-quark 
production at threshold and the high-energy limit. 
The asymptotic expressions of these two regimes are matched using the factorized form of the massive 
Wilson coefficients expressed in terms of massless coefficient functions and the massive operator matrix 
elements (OMEs), which are valid at momentum transfers $Q^2 \gg m_h^2 $, where $m_h$ denotes heavy quark 
mass~\cite{Behring:2014eya,Ablinger:2014nga,Ablinger:2016kgz,Ablinger:2017ptf}.
At NNLO one needs for this purpose the 3-loop OMEs, 
which are known exactly in part 
\cite{Ablinger:2010ty,Behring:2014eya,Ablinger:2014nga,Ablinger:2014vwa,Ablinger:2014lka,Ablinger:2017ptf} 
and are available in main terms in form 
of an approximation~\cite{Kawamura:2012cr} 
based on the fixed number of Mellin moments, calculated in Ref.~\cite{Bierenbaum:2009mv}.
Such approximations are commonly less accurate at small $x$, 
however their uncertainty can be validated using exact results, e.g., 
the recently calculated pure-singlet OME~\cite{Ablinger:2014nga}.
It turns out, that the exact pure-singlet term is well within the
uncertainties quoted for the approximate form obtained earlier 
from the first five non-vanishing Mellin moments~\cite{Bierenbaum:2009mv}, 
see Fig.~\ref{fig:aQj30}. Moreover, the exact pure-singlet term can be
employed to derive the gluon OME using the 
Casimir-scaling approximation. 
The expressions for the NNLO massive Wilson coefficients in the FFN scheme 
comprise all these ingredients~\footnote{The 3-loop massive OMEs 
obtained in this way can be also used to compute NNLO PDFs in the VFN scheme, see e.g. Ref.~\cite{Behring:2014eya}.}.

An important improvement in this formalism concerns definition of the heavy-quark mass. While the 
perturbative calculations are usually based on the pole mass-scheme, one rather turns to the  
{$\overline{\mathrm{MS}}\, $} running-mass for reasons of perturbative 
stability~\cite{Alekhin:2010sv}. 
Good agreement with the existing data is achieved by using 
this framework~\footnote{See Refs.~\cite{Ablinger:2014vwa,Ablinger:2014nga,Ablinger:2014lka}
for the scheme transformation to the $\overline{\rm MS}$ scheme up to 3-loops.}.

\begin{wrapfigure}{r}{0.5\textwidth}
  \centering
  \includegraphics[width=0.48\textwidth,height=0.45\textwidth]{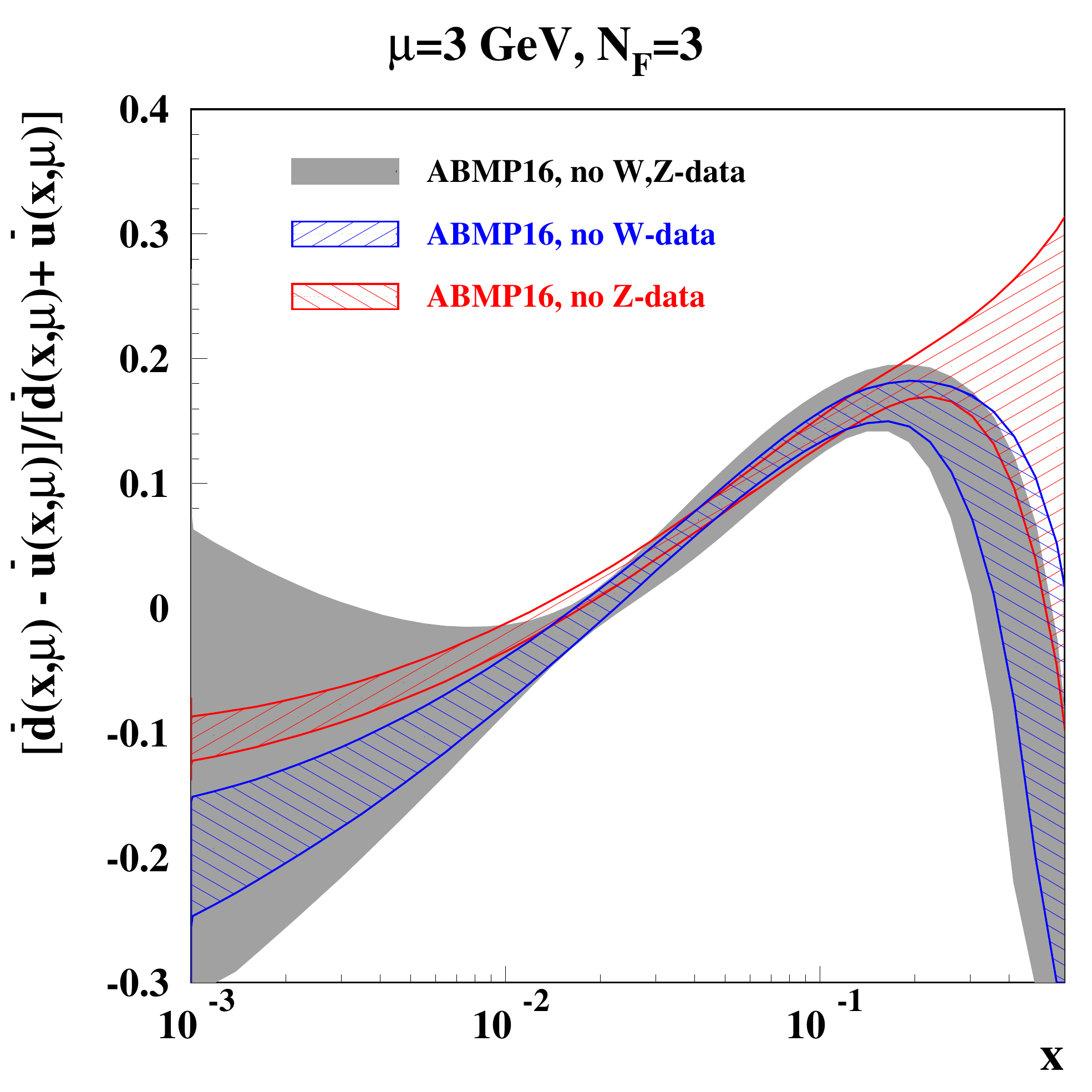}
  \caption{\small
The $1\sigma$ band for the NNLO quark iso-spin asymmetry 
$(\bar{d}-\bar{u})/(\bar{d}+\bar{u})$  in the 3-flavor scheme at the scale 
of $\mu=3~{\rm GeV}$ as a function of Bjorken $x$ obtained in variants of the ABMP16 PDF
fit~\protect\cite{Alekhin:2017kpj} with the data on production of $W$-bosons (left-titled hash), 
$Z$-bosons (right-titled hash), and both $W$- and $Z$-boson (shaded area) 
excluded form the fit. 
}
    \label{fig:wz}
\end{wrapfigure}
The value for the $c$-quark {$\overline{\mathrm{MS}}\, $} mass 
obtained in the recent ABMP16 fit~\cite{Alekhin:2017kpj}
$$
m_c(m_c)=1.252\pm 0.018 {\rm (exp.)}\pm 0.010 {\rm (th.)}
$$
is in a very good agreement with other precision determinations, e.g. based on the $e^+e^-$ data
\cite{Chetyrkin:2017lif}. 

The inclusive DIS data have a limited potential to disentangle the distributions of the quark 
species, particularly at small $x$. This is due to the fact that the HERA data consist only of proton data. 
Meanwhile, however, the Drell-Yan (DY) data from the LHC are of sufficient quality to determine the different flavor
distributions very well up to energies of 13~TeV.
These data probe the PDFs in a wide range of $x$, down to $x \sim 10^{-4}$ and
provide a variety of constraints on the quark distributions due to the production of both, $W^\pm$- and $Z$-bosons. 
The impact of this input on the PDF determination is demonstrated for instance
in ABMP16 fit~\cite{Alekhin:2017kpj}, which includes
a wide collection of the $W^\pm$- and $Z$-production data from the ATLAS, 
CMS and LHCb experiments at the LHC and from the D{\O} experiment at Tevatron. 
By discarding these data sets in a test variant of the ABMP16 fit we find
an essential deterioration in the determination of the quark distributions, 
leading to a greatly expanded uncertainty in the iso-spin asymmetry 
$(\bar{d}-\bar{u})/(\bar{d}+\bar{u})$ at small $x$. In the absence 
\begin{figure}[h]
  \centering
  \includegraphics[width=0.95\textwidth]{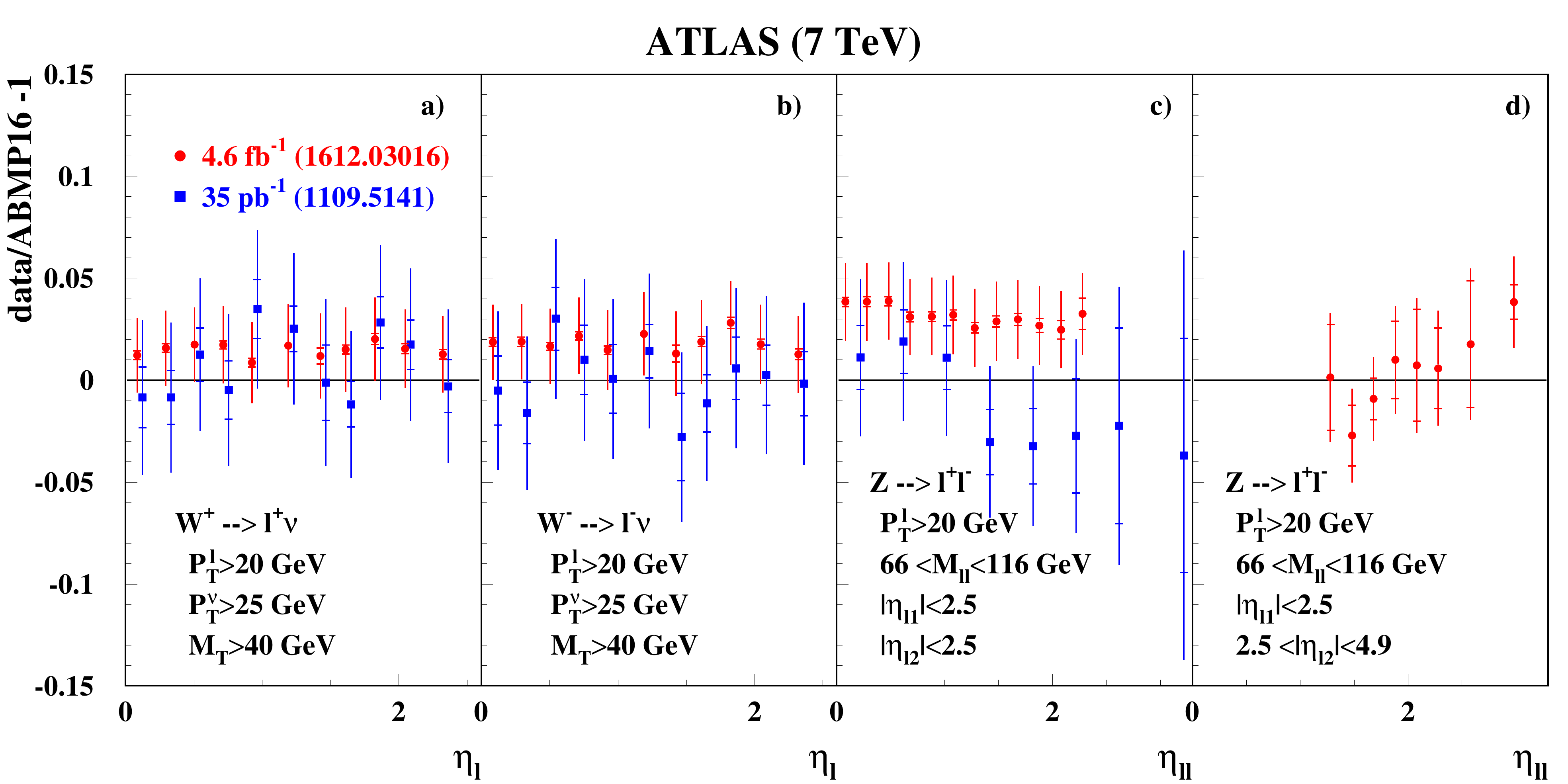}
  \caption{\small
    The pulls for the ATLAS data on the 
    $pp \to W^\pm+X \to l^\pm \nu + X$ production (a) and (b)
    and $pp \to Z+X \to l^+l^- + X$ (c): central region, (d): forward region) at $\sqrt s = 7$~TeV 
    collected at luminosity of 35 pb$^{-1}$ (2011)~\cite{Aad:2011dm} (blue squares)
    and 4.6 fb$^{-1}$ (2016)~\cite{Aaboud:2016btc} (red circles)
    with cuts on the lepton's transverse momentum $P_T^l>20~{\rm GeV}$ 
    as a function of the lepton pseudo-rapidity $\eta$ versus NNLO predictions
    obtained using {\tt FEWZ} (version 3.1)~\cite{Li:2012wna,Gavin:2012sy} and the ABMP16 PDFs.
}
    \label{fig:wzatlas}
\end{figure}
of DY data this piece is essentially unconstrained, see Fig.~\ref{fig:wz}. Therefore, in earlier PDF parameterizations, 
it was commonly set to zero for $x\to 0$. The collider DY data prefer a sizable negative value at 
$x\sim 10^{-4}$ and a symmetric non-strange sea is observed at $x\lesssim 10^{-5}$ only~\cite{Alekhin:2015cza}.

In general, the available DY data a very consistent. However, 
with rising experimental accuracy some tension between different experiments 
or even within one experiment may emerge. In particular, this concerns the recent 
ATLAS data on $W^\pm$- and $Z$-production at a center-of-mass (c.m.s.) energy of 7~TeV~\cite{Aaboud:2016btc}. 
This sample is in good agreement with the 
earlier data obtained by the same collaboration from the low-luminosity
run~\cite{Aad:2011dm}. It is in part related to the $W^\pm$-production, 
see Fig.~\ref{fig:wzatlas}. Meanwhile, the $Z$-production cross sections 
at central rapidity moved somewhat higher than the earlier ones. The tension is 
at the level of 1-2$\sigma$. It makes it difficult to describe 
the recent ATLAS data with the PDFs tuned to the previous release. 
Moreover, the 
epWZ16 PDFs extracted by ATLAS from data of Ref.~\cite{Aaboud:2016btc}, 
in combination with the inclusive DIS sample 
from HERA, demonstrate some unusual features, namely  
the strange sea is greatly enhanced 
if compared to strange suppression factors of $\sim 0.5$ as commonly 
obtained in the PDF fits. 

\begin{wrapfigure}{l}{0.45\textwidth}
  \centering
  \includegraphics[width=0.45\textwidth]{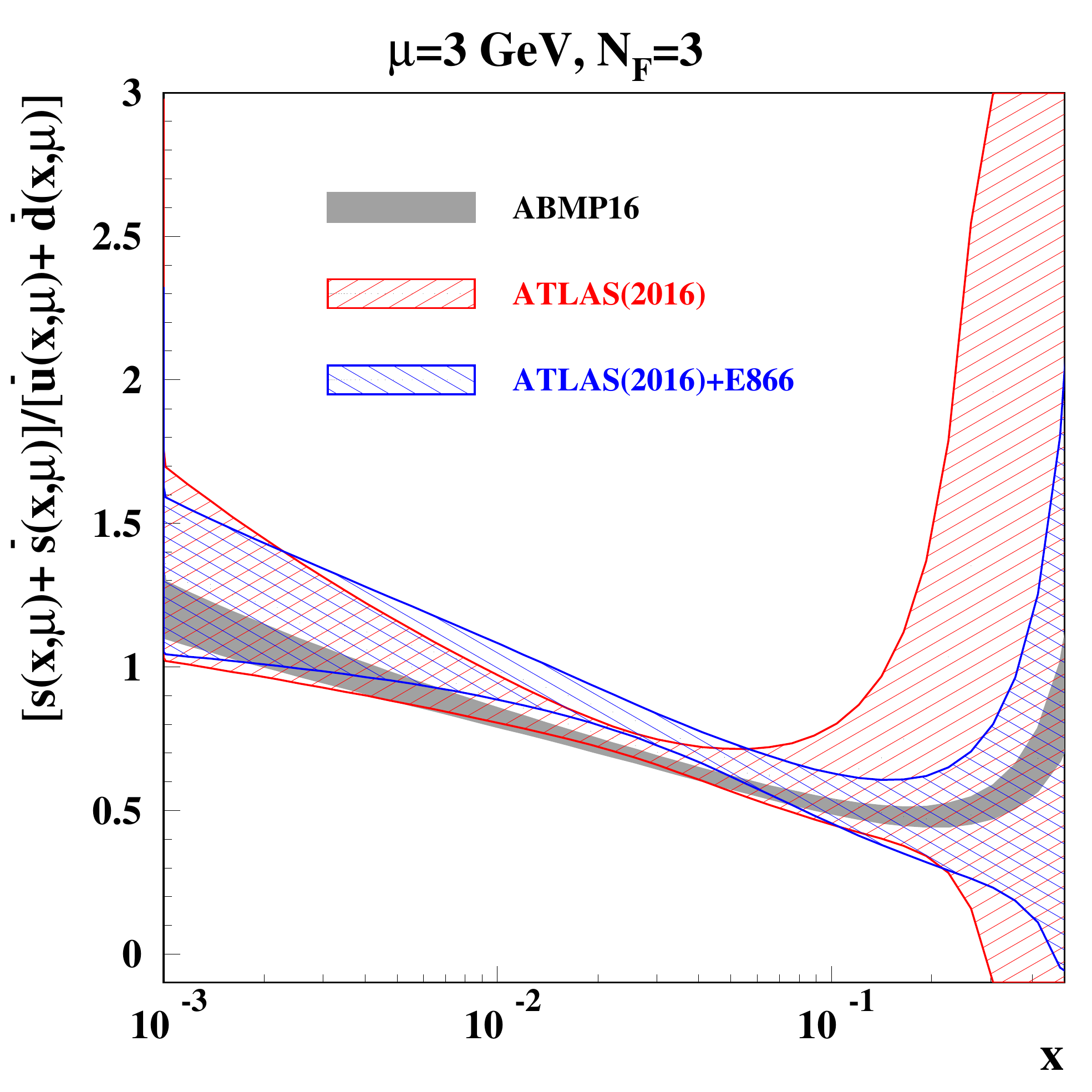}
  \caption{\small The same as in Fig.~\protect\ref{fig:wz}
for the strange sea suppression factor $(s+\bar{s})/(\bar{d}+\bar{u})$
obtained in test variants of the ABMP16 fit with ATLAS data used in 
combination with the inclusive HERA data (left-titled hash) and the E-866
data on the top (right-tilted hash) in comparison with the nominal ABMP16 PDFs
(shaded area).
}  
  \label{fig:ssupatlas}
%
\vspace*{6mm}
%
  \centering
  \includegraphics[width=0.45\textwidth]{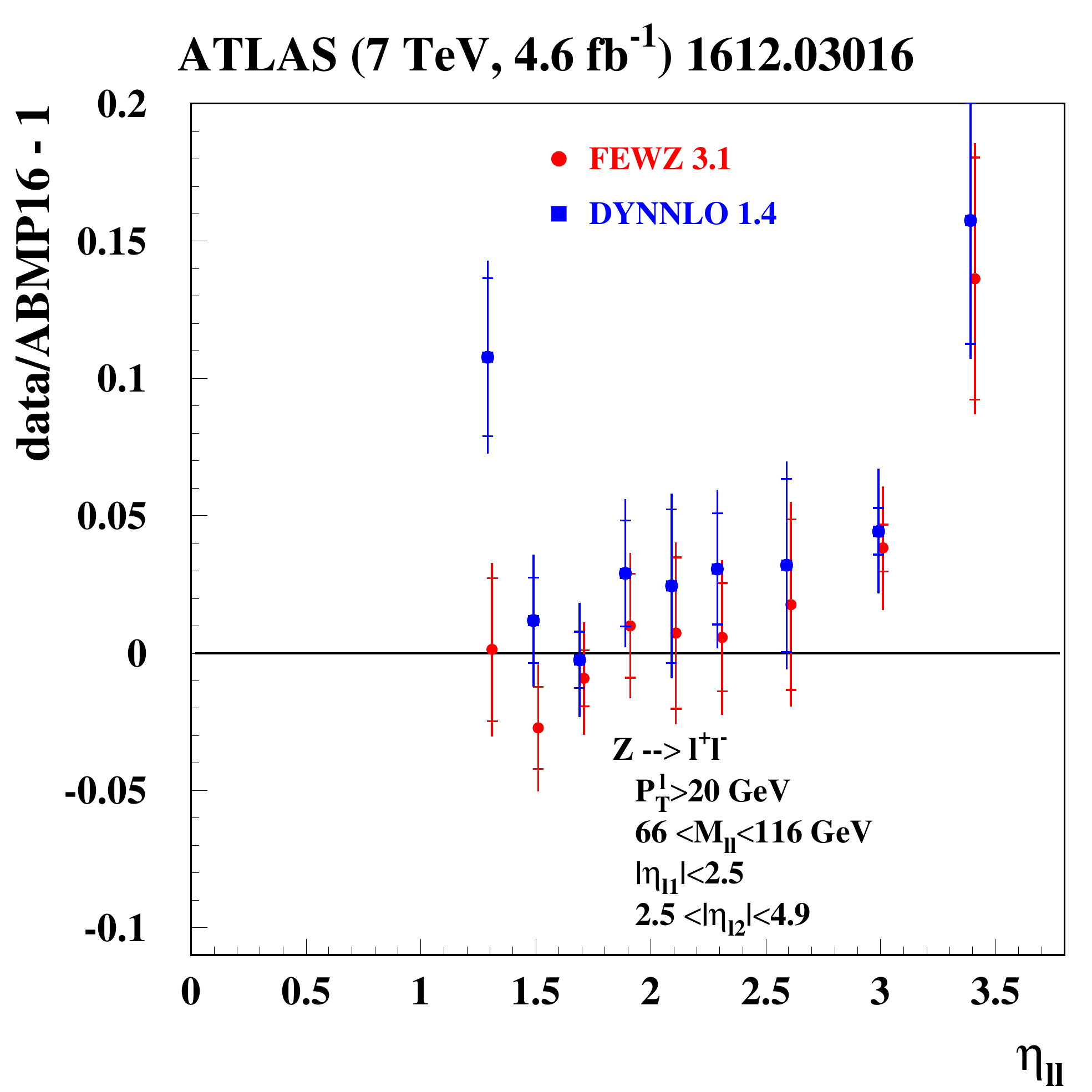}
  \caption{\small The same as in Fig.~\protect\ref{fig:wzatlas}
for the forward $Z$-productions data and the NNLO predictions 
obtained with {\tt FEWZ} (version 3.1)~\cite{Li:2012wna,Gavin:2012sy} 
(red circles) and {\tt DYNNLO} (version 1.4)
~\cite{Catani:2009sm,Catani:2007vq} (blue squares).}  
  \label{fig:dynnlo}
\end{wrapfigure}

To the most extent such an enhancement can be 
explained by a particular PDF shape employed in the analysis of 
Ref.~\cite{Aaboud:2016btc}. 
This shape had been suggested for the 
HERAPDF fit based on the HERA data only long ago. Therefore it contains many 
constraints due to the limited potential of inclusive DIS in disentangling 
quark distributions. 
By applying these constraints
the non-strange sea distributions are artificially suppressed and this 
suppression is compensated in the ATLAS analysis by the strangeness 
enhancement, which finally leads to
an abnormal strange sea suppression factor~\cite{Alekhin:2017olj}. 
If instead a flexible enough PDF shape is used, the strangeness preferred 
by the ATLAS data is in a reasonable agreement 
with the earlier determinations, although 
some tension at $x\sim 0.01$ still persists, see Fig.~\ref{fig:ssupatlas}.
This tension is evidently related to the impact of the upward shift in the 
central $Z$-production observed for the recent ATLAS measurements, 
see Fig.~\ref{fig:wzatlas}. However, it is worth noting that the 
ATLAS data for forward-rapidity demonstrate a different trend, although being  
statistically less significant. 

Besides, the CMS data on $Z$-production are also
somewhat lower than the ATLAS results, see Ref.~\cite{Alekhin:2017olj}
for details. Therefore this tension still deserves further clarification. 
Another problematic aspect of the DY data analysis concerns the accuracy of the 
tools, which are needed for the computation of the cross sections with account 
of realistic experimental cuts on the lepton transverse momentum.

The fully exclusive NNLO codes {\tt FEWZ}~\cite{Li:2012wna,Gavin:2012sy} and 
{\tt DYNNLO}~\cite{Catani:2009sm,Catani:2007vq},
which accomplish these 
computations, are not in 
perfect agreement in the relevant kinematical region, see Fig.~\ref{fig:dynnlo}.   
In general, the predictions by {\tt DYNNLO} are lower than the ones by {\tt FEWZ} by $\sim 1\%$. 
However, at the edges of the distributions this difference rises to 10\%.
The discrepancies between {\tt DYNNLO} and {\tt FEWZ} were partially understood as being due
to the numerical integration accuracy~\cite{Yannick} and due to effects of 
experimental cuts on the lepton transverse momentum in higher-order QCD computations~\cite{Frixione:1997ks}, 
but at the moment the theoretical accuracy is limiting the related studies~\footnote{In the ATLAS 
analysis~\cite{Aaboud:2016btc} the {\tt DYNNLO} calculations are used for nominal 
results and the difference between {\tt DYNNLO} and {\tt FEWZ} is taken as a
theoretical uncertainty.}.

\begin{wrapfigure}{l}{0.5\textwidth}
  \centering
  \includegraphics[width=0.45\textwidth]{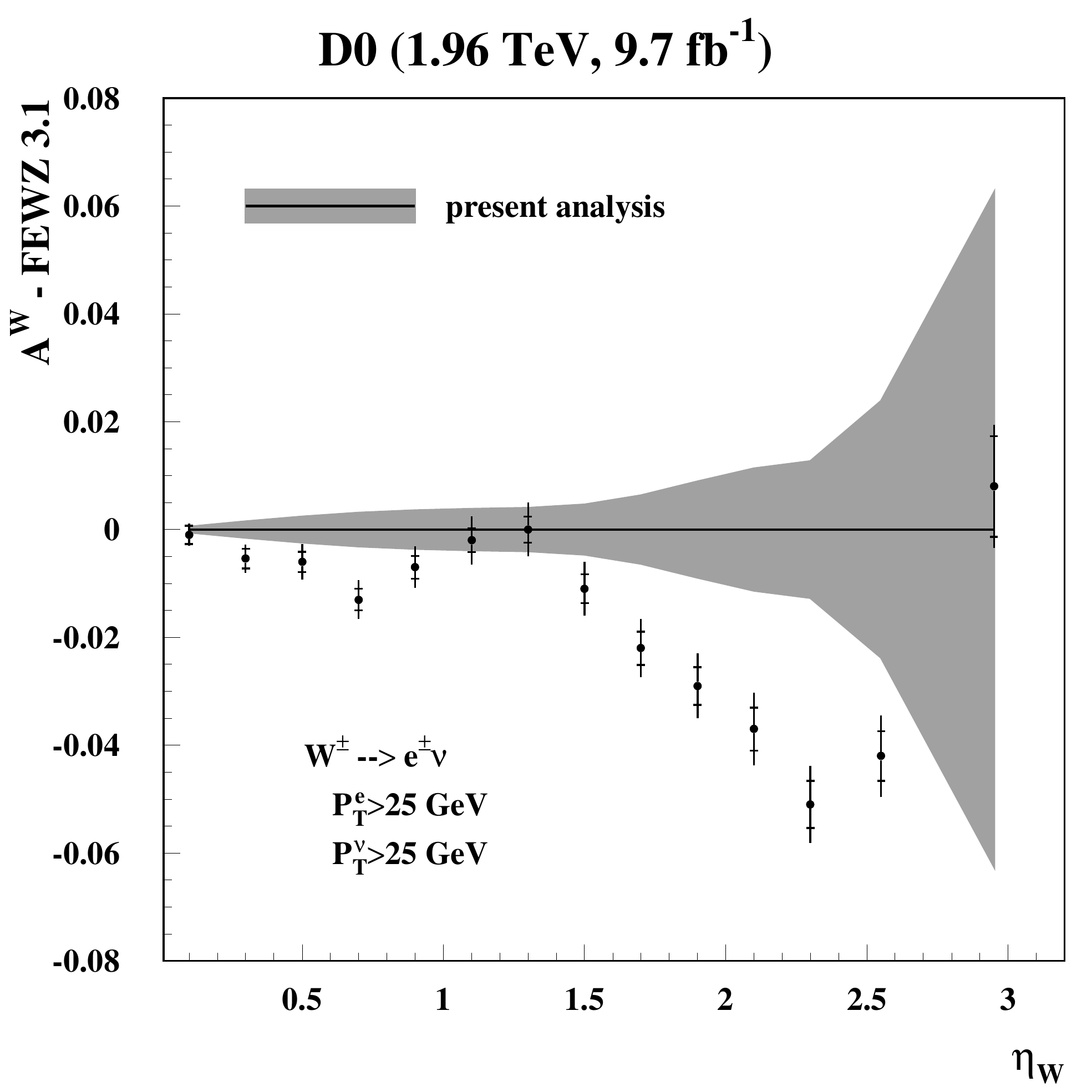}
  \caption{\small The pulls of the D\O~ data on the $W$ 
charge asymmetry~\protect\cite{Abazov:2013dsa} versus the predictions obtained
with {\tt FEWZ} (version 3.1) at NNLO in QCD and the ABMP15 
PDFs~\protect\cite{Alekhin:2015cza} constrained by the D\O~ 
$e$-asymmetry data~\protect\cite{D0:2014kma}
as a function of the $W$-boson rapidity $\eta_W$.
The shaded area displays the PDF uncertainties in the predictions. }
  \label{fig:d0w}
\end{wrapfigure}

The DY collider data also help to constrain
the large-$x$ region of the quark distributions, in particular for the 
ratio $d/u$. In this context the 
D{\O} measurement of the $W$ charge asymmetry~\cite{Abazov:2013dsa} 
provides the statistically most significant constraint. 
Since $W$-boson production is not measured directly,  
the $W$-asymmetry is derived in the D{\O} analysis 
from the measurement of the electrons stemming from the $W$ decays. 
This is possible in a unique way at leading order (LO) only, while 
account of the higher-order corrections requires additional modeling. This, in
particular, causes sensitivity to the $W$-asymmetry obtained by the choice 
of the PDFs used. It leads to a certain tension between 
the $W$-asymmetry data and the original $e$-asymmetry ones, if the PDFs 
are varied. 
In particular, the predictions of the $W$-asymmetry for the 
D{\O} kinematics obtained with the ABMP15 PDFs based on the D{\O}
data on the $e$-asymmetry~\cite{D0:2014kma}, 
are in substantial disagreement with the D{\O} data 
on $W$-asymmetry, see Fig.~\ref{fig:d0w}.

\begin{figure}
  \centering
  \includegraphics[width=0.45\textwidth]{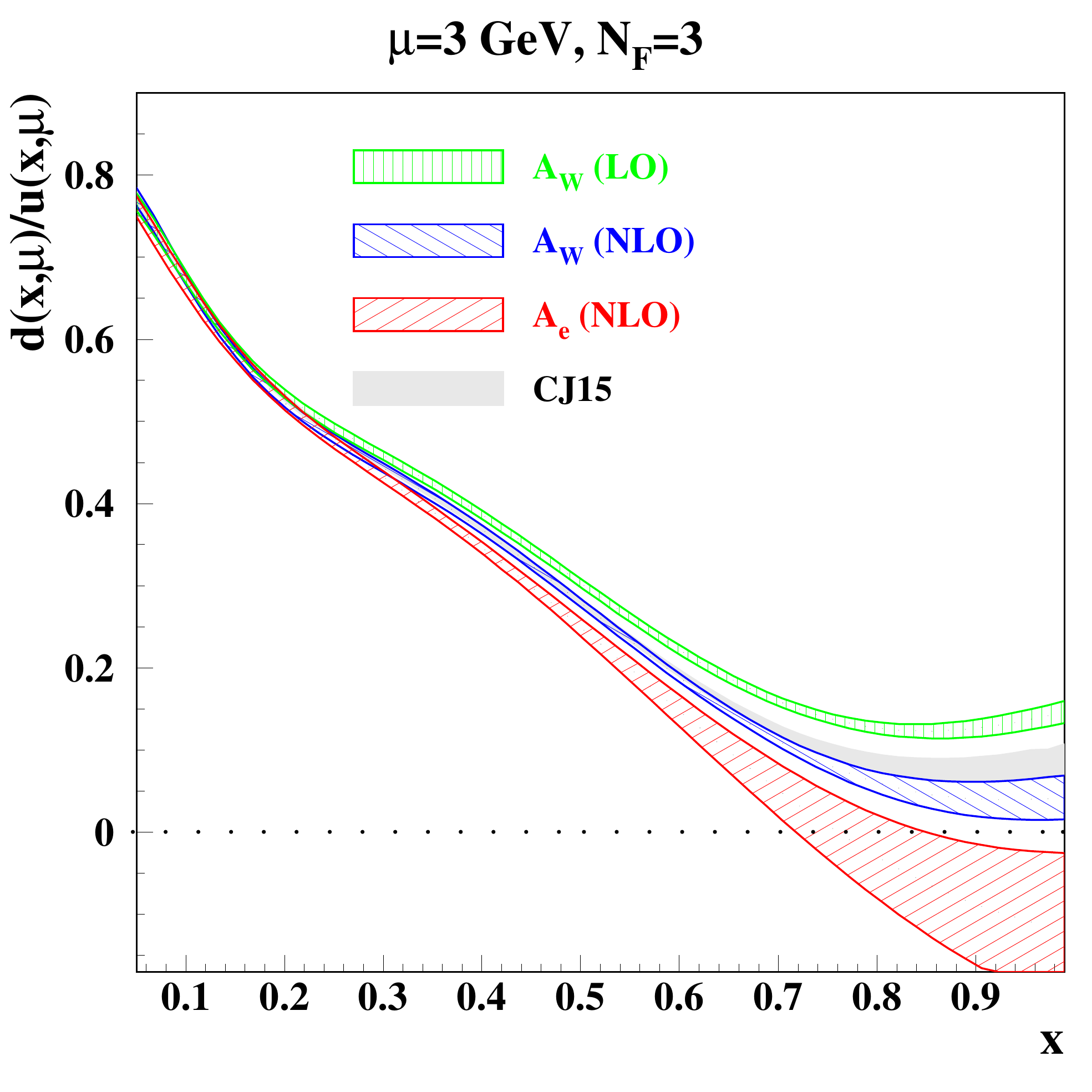}
  \includegraphics[width=0.45\textwidth]{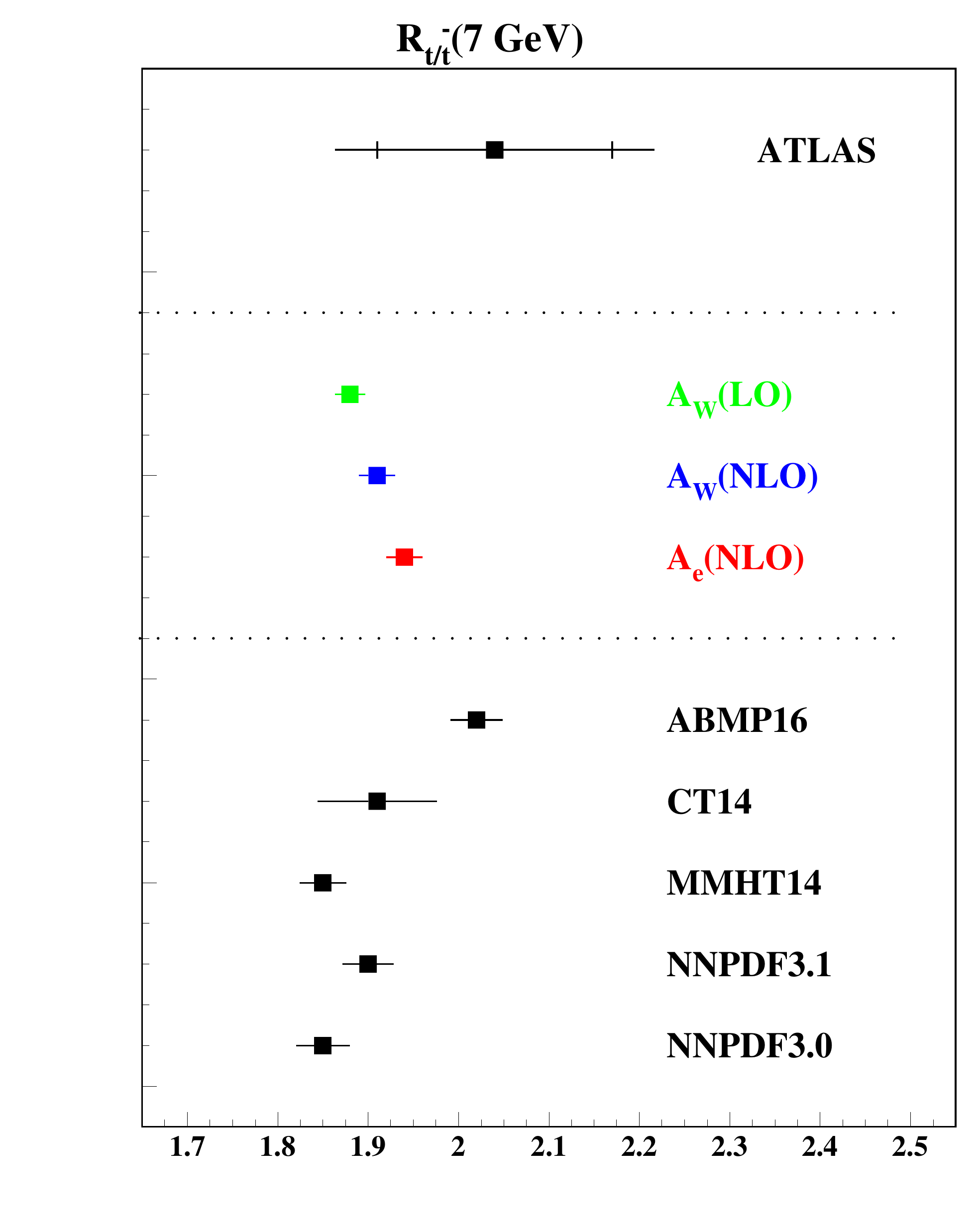}
  \caption{\small Left: The same as in Fig.~\protect\ref{fig:wz}
for the ratio $d/u$ obtained using the CJ15 PDF 
shape~\protect\cite{Accardi:2016qay} and 
with addition of the D{\O} data on 
$W$- and $e$-asymmetry, described within various approximations
(vertical hash: $W$-asymmetry~\protect\cite{Abazov:2013dsa} at LO, 
left-titled hash: the same at NLO, 
right-tilted hash: $e$-asymmetry~\protect\cite{D0:2014kma} at NLO
in comparison with the nominal CJ15 PDFs
(shaded area). Right: The ratio of single top to anti-top production cross 
section in $pp$ collisions at c.m.s. energy 7 TeV computed with the PDFs 
obtained in these variants of the fit in comparison with the ATLAS 
data~\protect\cite{Aad:2014fwa} and the predictions of 
ABMP16~\protect\cite{Alekhin:2015cza}, CT14~\protect\cite{{Dulat:2015mca}}, 
MMHT14~\protect\cite{Harland-Lang:2014zoa}, 
NNPDF3.0~\protect\cite{Ball:2014uwa} and 
NNPDF3.1~\protect\cite{Ball:2017nwa} PDFs.
}  
  \label{fig:cj15}
\end{figure}

The potential of the D{\O} measurements on the large-$x$ asymptotics of the
$d/u$ ratio was checked in the recent CJ15 PDF fit~\cite{Accardi:2016qay}. 
An advantage of this analysis is a flexible PDF shape, which allows 
for a non-vanishing value of $(d/u)\vert _{x=1}$. The CJ15 analysis combines 
both the $W$-asymmetry and the $e$-asymmetry D{\O} data. 
The large-$x$ $d/u$ ratio is mainly driven by the $W$-asymmetry
data due to its statistical significance. The impact of these 
data is quite sensitive on the theoretical accuracy of the analysis.
The $d/u$ ratio obtained with the LO description leads to higher values 
than the one obtained accounting for the next-to-leading order (NLO) corrections, 
see Fig.~\ref{fig:cj15}. 
Furthermore, the uncertainties in the $d/u$-ratio 
do substantially rise in the NLO fit. This is evidently due to the smearing of the predictions 
by the gluon-initiated contribution and the propagation of the uncertainty 
of the gluon-distribution into the ratio of $d/u$ extracted from the fit. 
The theory framework of the CJ15 fit is based on a $K$-factor approximation 
of the $W$-production cross section, with the NLO predictions represented as a product of 
the LO approximation and the pre-computed ratio of the NLO and LO cross sections. In case
of the $\bar{p}p$ initial state such an approach reproduces the initial LO predictions.
Therefore the CJ15 result on $d/u$ should be biased upwards due to the missing 
NLO corrections, see Ref.~\cite{Alekhin:2018dbs} for details. 
The value of $d/u$ preferred by the  D{\O} data on 
the $e$-asymmetry~\cite{D0:2014kma} is substantially lower than the 
$W$-asymmetry results and even extends to negative values at $x\to 1$, 
although with large uncertainties, see Fig.~\ref{fig:cj15}. 

Comparing it with 
the NLO determination based on the $W$-asymmetry, we conclude that there is no strong 
evidence in favor of a non-vanishing $(d/u)\vert _{x=1}$ from the analysis of the 
 D{\O} data. Moreover, the
$e$-asymmetry data, preferring a smaller value of $d/u$, are less model-dependent 
than the $W$-asymmetry.

The interpretation of the D{\O} data in the PDF fit turns out to be essential
for the related phenomenology of electroweak single-top  
production since the latter reaction is to a great extent driven by the quark-initiated 
subprocesses. Therefore a trend observed for the $d/u$ ratio    
in the variants of the PDF fit with a different treatment of the  D{\O} experimental 
input is reflected in the ratio of the top and anti-top production cross sections
$R_{t/\bar{t}}$
computed with respective PDFs, see Fig.~\ref{fig:cj15}. 
For the fit based on the $e$-asymmetry data the value of $R_{t/\bar{t}}$ is larger by $\sim 2 \sigma$ than 
for the one obtained from the LO fit using the $W$-asymmetry data.   
This is comparable to the spread in the predictions of different PDFs, which can 
be explained in part by the selection of the DY collider data and their treatment. 

In summary, we have considered some current trends, achievements and problems in the precision 
determination of PDFs. For the DIS data a solid theoretical framework is available  
with the FFN scheme used for description of the heavy-quark contribution.
It provides good agreement with 
existing experimental data in a wide range of momentum transfers and implies
a consistent setting of the heavy-quark masses, which are basic parameters of the Standard Model.
The emerging DY data 
collected at the hadron colliders start to play a crucial role in disentangling 
quark species, particularly at small and large values of $x$. These 
new inputs demonstrate a good overall consistency with the earlier constraints 
on PDFs coming from the fixed-target experiments. No 
dramatic change in the PDFs is caused in case of consistent account of the 
higher-order QCD corrections and using PDF shapes which are flexible enough
for fitting the experimental data.   

%


\end{document}